\documentclass[a4paper,11pt]{article}
\usepackage{pos}

\title{GAN-based data augmentation for rare and exotic hadron searches in Pb--Pb collisions in ALICE}

\author[a]{Anisa Khatun on behalf of the ALICE Collaboration}

\affiliation[a]{University of Foggia and INFN\\
Foggia, Italy}

\emailAdd{anisa.khatun@cern.ch}

\abstract{This work presents a feasibility study aimed at enhancing the reconstruction sensitivity for rare heavy-flavour hadrons in Pb–Pb collisions in the ALICE experiment, using the $\Xi_{\mathrm{c}}^{+}$ baryon as a benchmark. 
The $\Xi_{\mathrm{c}}^{+}$ baryon has a low rate of production and some complex decay topologies as for instance the decay $\Xi_{\mathrm{c}}^{+} \rightarrow \Xi^{-} + \pi^{+} + \pi^{+}$ considered in this work. Traditional simulation workflows involving event embedding and full detector response are computationally expensive and statistically limited, especially for rare signals.
This study represents the first exploration of generative models within the heavy-flavour programme of ALICE. 
It uses a dataset of reconstructed physics quantities, such as momenta, positions, and decay vertex coordinates of $\Xi_{\mathrm{c}}^{+}$ decay products in Pb–Pb collisions as input features, derived from augmented ALICE Monte Carlo simulations.
Such features will serve as a training set for Generative Adversarial Networks (GANs) designed to generate statistically significant synthetic signal samples without the need for additional full simulations. While $\Xi_{\mathrm{c}}^{+}$ serves as a benchmark, the broader objective is to enable searches for exotic heavy-flavour hadrons or other exotic states with complex decay patterns.
By leveraging GAN-based augmentation, this approach supports rare-signal extraction in computationally demanding analyses and opens the way to broader applications of generative models in the ALICE heavy-flavour programme.
}

\FullConference{
}

\begin{document}
\maketitle

\section{Introduction}
The study of heavy-flavour and exotic hadrons in ultra-relativistic heavy-ion
collisions provides essential insight into the properties of the Quark--Gluon
Plasma (QGP). However, searches for rare and short-lived states are often limited
by low production rates and by the large combinatorial background inherent to
high-multiplicity Pb--Pb collisions.
In the ALICE experiment, standard Monte Carlo (MC) simulation workflows of heavy-ion collisions rely
on event embedding and full detector response, which are computationally
expensive and statistically constrained for rare signals.

In these proceedings, we explore the feasibility of using Generative Adversarial
Networks (GANs) as a data augmentation tool to enhance the statistical reach of
rare heavy-flavour hadron analyses. The approach aims to generate synthetic
samples of reconstructed physics observables that reproduce the distributions
and correlations of MC-generated signal candidates, without requiring additional
full detector simulations.

\section{Benchmark physics case: \texorpdfstring{$\Xi_{\mathrm{c}}^{+}$}{Xi_c+} baryon}
\begin{figure}[h!]
    \centering
    \includegraphics[width=0.45\linewidth]{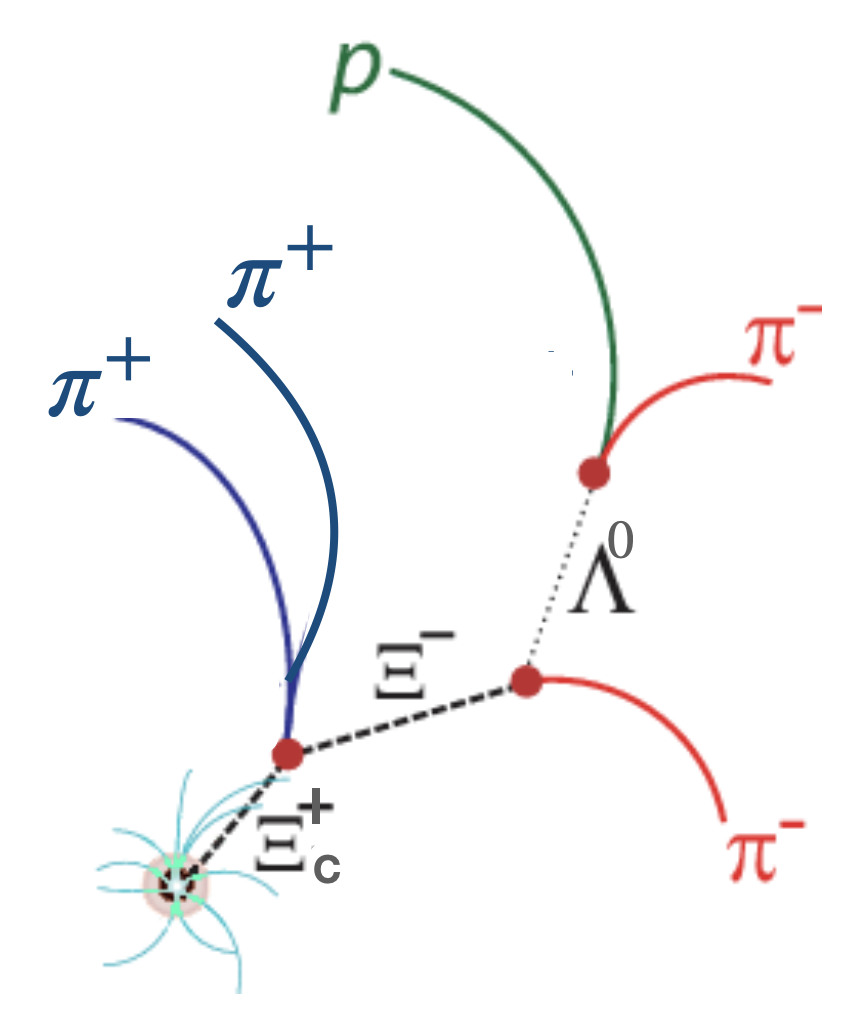}
    \caption{$\Xi_{\mathrm{c}}^{+}$ decay chain.}
    \label{fig:decaytopology}
\end{figure}

The $\Xi_{\mathrm{c}}^{+}$ baryon is chosen as a benchmark due to its rare
production and complex decay topology.
In this study, the decay channel
$\Xi_{\mathrm{c}}^{+} \rightarrow \Xi^{-} + \pi^{+} + \pi^{+}$ is considered,
which involves a cascade decay with multiple secondary vertices as shown in Fig.~\ref{fig:decaytopology}.
Such topologies pose significant reconstruction challenges in Pb--Pb collisions,
where track density and background levels are high.

While the $\Xi_{\mathrm{c}}^{+}$ baryon serves as a reference case, the methodology
presented here is designed to be generic and applicable to searches for other
rare or exotic heavy-flavour states with similarly complex decay patterns~\cite{xicRun2pp}.

\section{GAN-based data augmentation strategy}
Generative Adversarial Networks are a class of machine learning models composed
of two competing neural networks: a generator and a discriminator.
The generator aims to produce synthetic data samples that resemble the training
data, while the discriminator attempts to distinguish between real and generated
samples. Through this adversarial process, the generator learns to model the underlying
data distribution~\cite{Goodfellow}.

\begin{figure} [h!]
    \centering
    \includegraphics[width=0.85\textwidth]{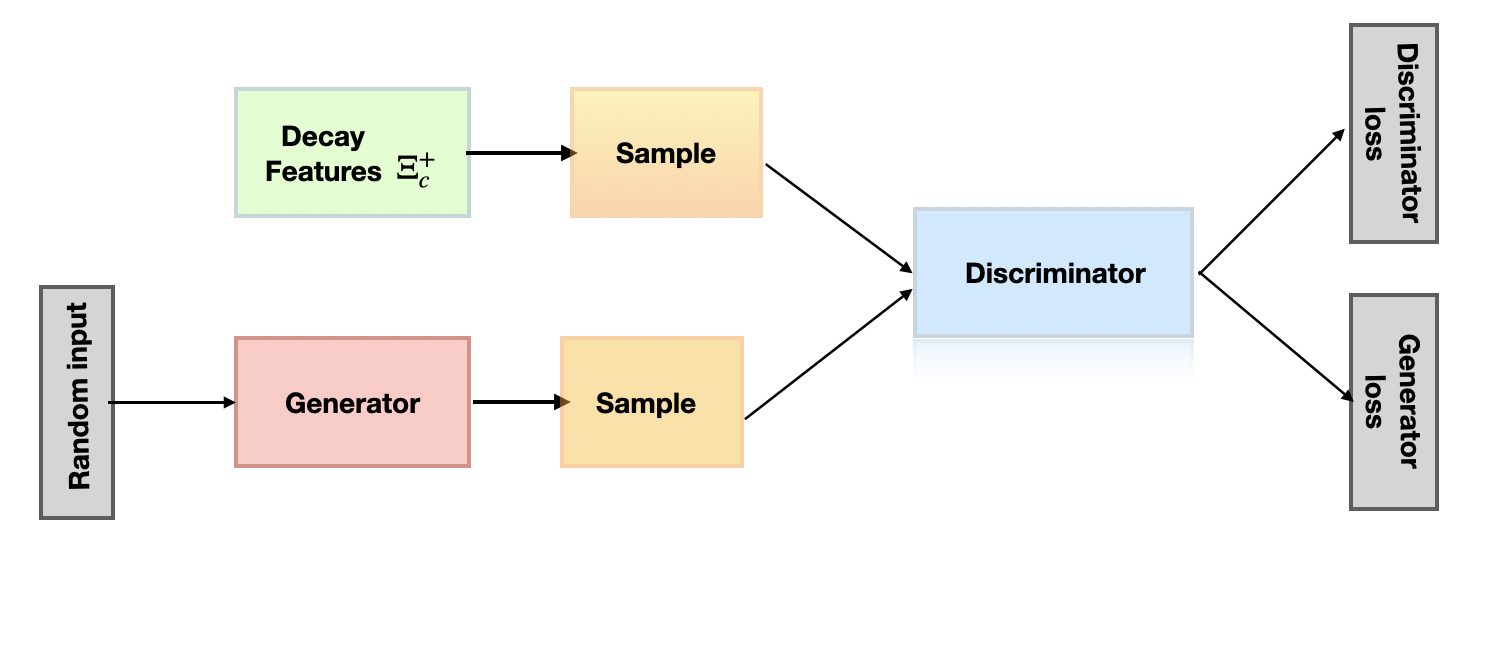}
    \caption{Schematic representation of the Generative Adversarial Network (GAN)
architecture used in this study. The generator produces synthetic reconstructed
features starting from random noise, while the discriminator attempts to
distinguish generated samples from real ALICE Monte Carlo data.}
 \label{fig:GANarchitech}
\end{figure}

In this work, the GAN is trained on reconstructed
topological and kinematic observables of candidate $\Xi_{\mathrm{c}}^{+}$ baryons decaying in the $\Xi_{\mathrm{c}}^{+} \rightarrow \Xi^{-} + \pi^{+} + \pi^{+}$ decay channel obtained from MC simulations.
The set of input feature includes variables such as decay lengths, pointing angles,
distances of closest approach (DCA) to the primary vertex, and kinematic
quantities of the decay products.
Once trained, as demonstrated in Fig.~\ref{fig:GANarchitech}, the GAN can produce statistically significant synthetic signal
samples that mimic the MC distributions and correlations of these observables.

\section{GAN training and validation}
\begin{figure}[h!]
    \centering
    \includegraphics[width=0.95\textwidth]{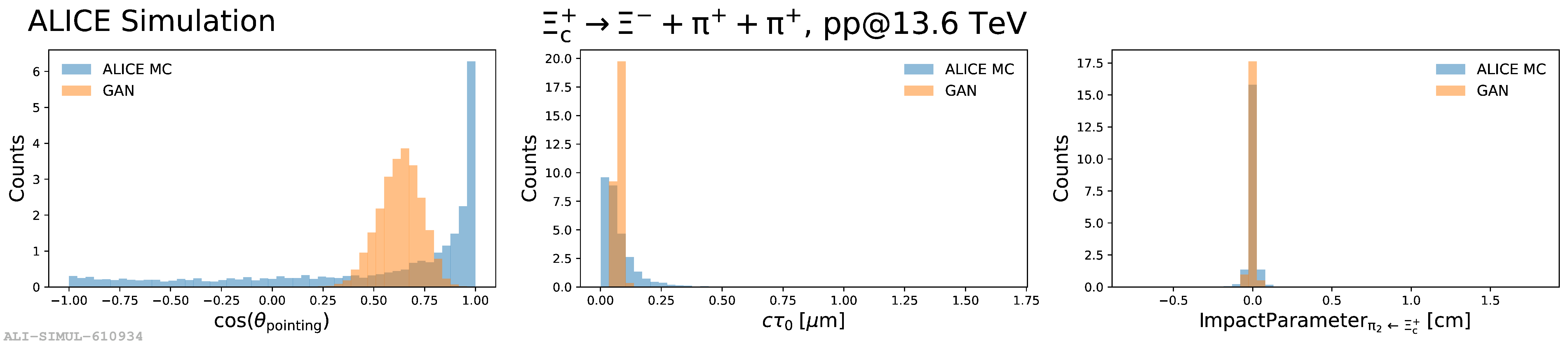}
    \caption{Comparison of reconstructed feature distributions between
GAN-generated samples and ALICE Monte Carlo at the beginning of the training.}
    \label{fig:epoch0}
\end{figure}

The GAN is trained using reconstructed $\Xi_{\mathrm{c}}^{+}$ signal candidates
obtained from ALICE MC simulations.
At early stages of the training, the generated feature distributions show
significant discrepancies with respect to the MC reference, as illustrated in
Fig.~\ref{fig:epoch0}.
This behavior is expected before the adversarial networks reach convergence.
With increasing training epochs, the agreement between GAN-generated samples and
MC improves, indicating stable adversarial learning.

The quality of the generated samples is assessed by comparing both
one-dimensional distributions and two-dimensional correlations between real
MC and GAN output.
Statistical compatibility is quantified using the Kolmogorov--Smirnov (KS) test,
which measures the maximum distance between the cumulative distribution functions
of two samples~\cite{KSoriginal,Stephens,PressNR}.
For each reconstructed observable, a KS test is performed between the ALICE MC
reference and the GAN-generated sample.

\begin{figure}[h!]
    \centering
    \includegraphics[width=0.95\textwidth]{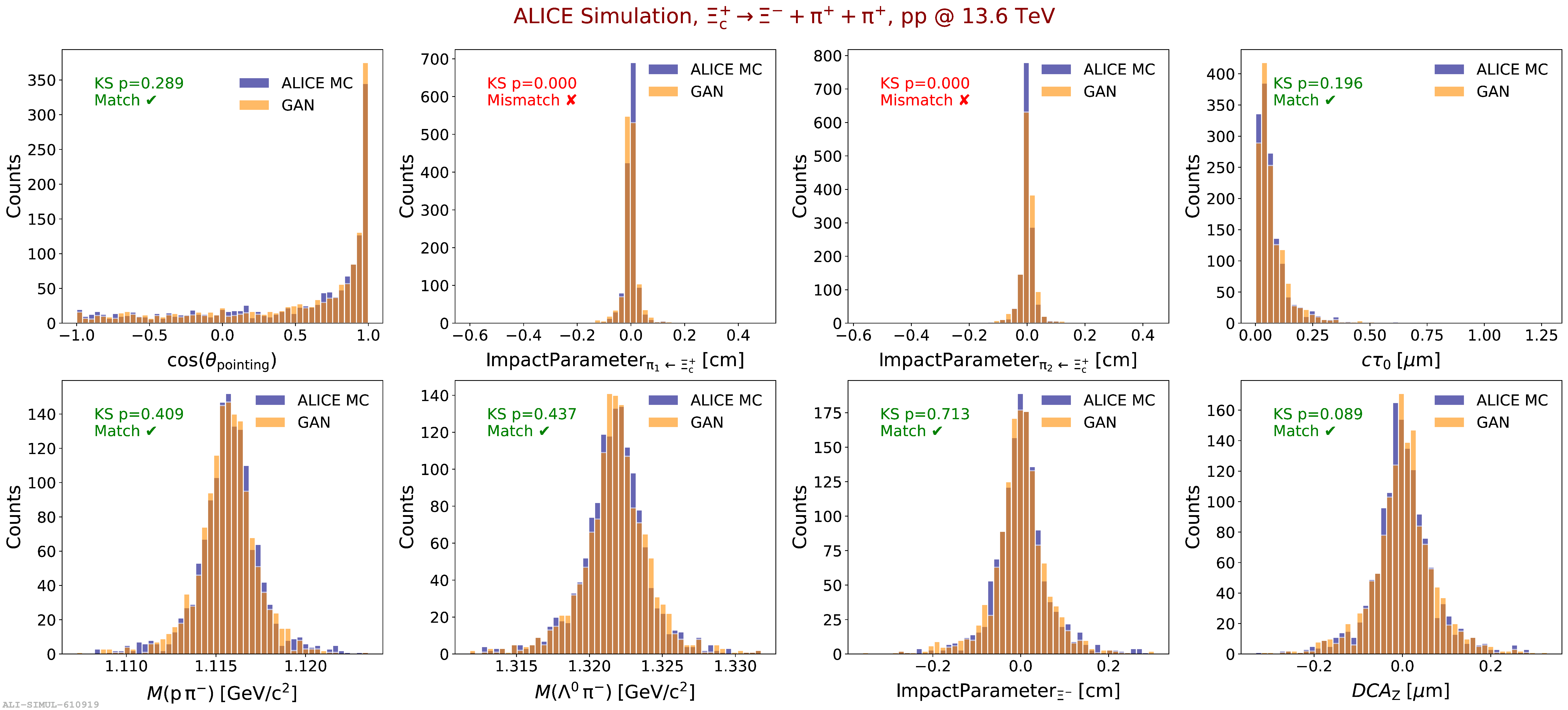}
    \caption{Comparison of one-dimensional reconstructed feature distributions
between GAN-generated samples and ALICE Monte Carlo after training.
The corresponding Kolmogorov--Smirnov p-values quantify the statistical
compatibility between the two samples for each observable.}
    \label{fig:GANeval}
\end{figure}

The resulting p-value represents the probability that the two samples are drawn
from the same underlying distribution.
Large p-values ($> 0.05$) indicate statistical compatibility, while small p-values ($<0.05$) signal significant discrepancies.
As shown in Fig.~\ref{fig:GANeval}, several observables exhibit p-values above
commonly used compatibility thresholds, demonstrating that the GAN is able to
reproduce the relevant physics distributions within statistical uncertainties.

Beyond reproducing individual feature distributions, preserving correlations
among variables is essential for realistic physics modelling.

\begin{figure}[h!]
    \centering
    \includegraphics[width=0.95\linewidth]{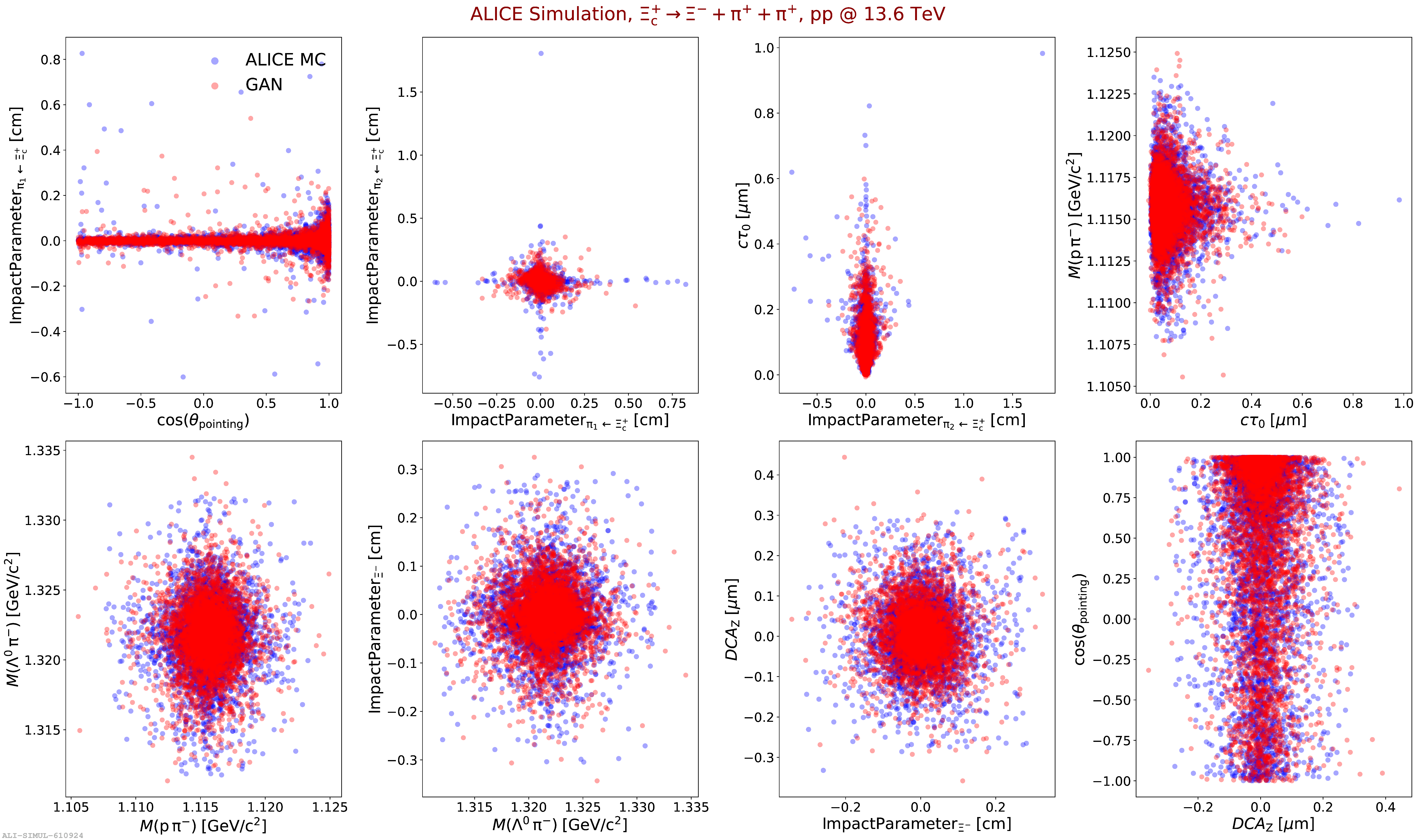}
    \caption{Two-dimensional scatter plots illustrating correlations between
selected reconstructed observables for GAN-generated samples and ALICE MC.}
    \label{fig:Corr}
\end{figure}

Figure~\ref{fig:Corr} presents two-dimensional scatter plots comparing correlations
between selected observables for ALICE MC and GAN-generated samples.
Despite a few outliers in some features, the close agreement observed in both shape and density demonstrates that the GAN
captures not only marginal distributions but also the underlying multi-dimensional
structure of the signal feature space.

\begin{figure}[h!]
    \centering
    \includegraphics[width=0.85\textwidth]{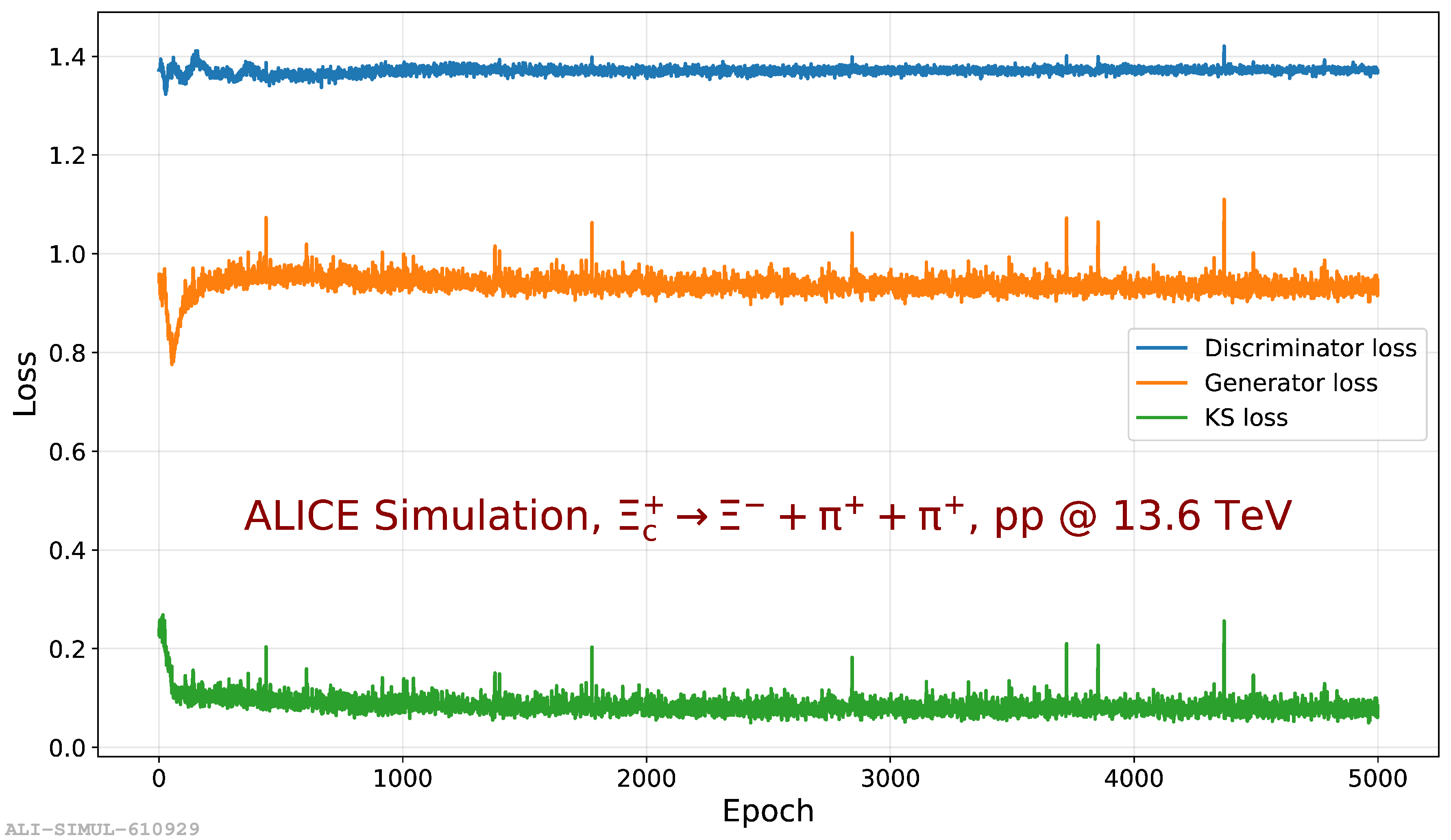}
    \caption{Evolution of the generator loss, discriminator loss, and KS-based
validation metric as a function of training epoch.}
    \label{fig:TrainingStability}
\end{figure}

The stability of the adversarial training is further evaluated by monitoring the
evolution of the generator loss, discriminator loss, and the KS-based validation
metric as a function of the training epoch.
As shown in Fig.~\ref{fig:TrainingStability}, the loss functions exhibit a stable
behavior over approximately $1.5\times10^{3}$ training epochs, indicating the
absence of mode collapse and confirming the robustness of the GAN training.

\section{Outlook for Pb--Pb analyses}
In the full Pb--Pb analysis workflow, GAN-augmented signal samples seem a promising approach to efficiently use computing resources for the training of machine learning classifiers and to test the feasibility
of rare signal extraction under realistic heavy-ion conditions.
The performance of such a new approach can be validated using standard metrics such as signal
significance, background rejection, and stability against analysis variations.

Future developments foresee extending the approach to a larger set of observables,
exploring more advanced GAN architectures, and adapting the training strategy to
the increased complexity of Pb--Pb collision environments at LHC energies.

\section{Conclusions}
 This study demonstrates the feasibility of GAN-based data augmentation within the heavy-flavour program of the ALICE experiment.
The results demonstrate that GANs can successfully reproduce reconstructed
physics observables and their correlations for rare heavy-flavour signals.
This approach offers a promising path to alleviate computational limitations and
to enhance sensitivity in searches for rare and exotic hadrons in heavy-ion
collisions.

\end{document}